\documentclass[11pt]{article}

\input{diagrams}
\usepackage{amsmath,amssymb}
\def\text#1{\mbox{\rm #1\ }}
\def\ie{{\rm i.e.,\/}\ }
\def\etc{{\rm etc.\/}\ }
\def\cf{{\rm cf.\/}\ }
\def\id{\mbox{\it id\,}}
\def\one{\mbox{\rm 1}\hskip-2.8pt \mbox{\rm l}}
\newcommand{\scal}[2]{\langle \; #1 \; ,\; #2 \; \rangle}
\newcommand{\cro}[1]{\lbrack \, #1 \, \rbrack}
\newcommand{\Mlo}[1]{\ensuremath{(M_{#1}(\Lambda^{2}))_0}}

\newcommand{\inplus}{\ensuremath{\,\,\subset\!\!\!\!\!\!+\,\,\,\,}}
\newcommand{\udl}[1]{\underline{#1}}
\def\ZZ{\mbox{\rm Z}\hskip-5pt \mbox{\rm Z}}

\def\CC{\mbox{\rm C}\hskip-5.5pt \mbox{l} \;}

\title{Action of a finite quantum group on the algebra of complex
$N\times N$ matrices
   \vspace{0.7cm} \\
}

\author{R. Coquereaux${}^1$\thanks{~Email: coque\@cpt.univ-mrs.fr},
        G.E. Schieber${}^1$\thanks{~Email: schieber\@cpt.univ-mrs.fr}\\
\\
${}^1$ {\it Centre de Physique Th\'eorique - CNRS - Luminy, Case 907} \\
       {\it F-13288 Marseille Cedex 9 - France} \\
\\
}

\date{July 11, 1998\thanks{Contribution to the International Conference
``Particles, Fields and Gravitation''
devoted to the memory of Professor Ryszard R{\c{a}}czka (1931-1996)
Lodz, April 15-19, 1998. Proceedings to be published by AIP.}}

\begin{document}

\thispagestyle{empty}
\begin{titlepage}

\maketitle

\vfill

\abstract{
Using the fact that the algebra ${\mathcal M} \doteq M_N(\CC)$ of $N \times
N$ complex matrices
can be considered as a reduced quantum plane, and that it is a
module
algebra for a finite dimensional Hopf algebra quotient ${\mathcal H}$ of
$U_{q}sl(2)$ when $q$ is
a root of unity, we reduce this algebra ${\mathcal M}$ of matrices
(assuming $N$
odd) into indecomposable modules for ${\mathcal H}$.
We also show how the same finite dimensional quantum group acts on
the space of generalized differential forms defined as the
reduced Wess Zumino complex associated with the algebra ${\mathcal M}$.
}

\vspace{1.2 cm}


\vspace{1.0cm}

\noindent Anonymous ftp or gopher: cpt.univ-mrs.fr

\vspace{0.5 cm}

\noindent {\tt math.QA/9807016}\\
\noindent CPT-98/P.3668

\vspace*{0.3 cm}

\end{titlepage}


\section {Introduction}

When $q$ is a root of unity ($q^{N}=1$), the quantized enveloping
algebra $U_{q}sl(2,\CC)$ posesses interesting quotients that are
finite dimensional Hopf algebras.
The structure of the left regular representation of such an algebra was
investigated  in  \cite{Alekseev} and the pairing with its dual in \cite{Gluschenkov}.
We call ${\mathcal H}$ the Hopf algebra quotient of
$U_{q}sl(2,\CC)$  defined by the
relations $K^{N}=1, X_{\pm}^{N}=0$ (we shall define the generators
$K, X_{\pm}$ in a later section), and ${\mathcal F}$ its dual.
It was shown\footnote{Warning: the authors 
of \cite{Alekseev} actually consider a Hopf algebra quotient 
defined by $K^{2N} = 1, X_{\pm}^{N}=0$, so that their algebra is, in 
a sense, twice bigger than ours.} in \cite{Alekseev}
that the {\underline{non}} semi-simple algebras ${\mathcal H}$ is isomorphic with the
direct sum of a complex
matrix algebra and of several copies of suitably defined matrix algebras
with coefficients in the ring $Gr(2)$ of Grassmann numbers with two generators.
The explicit structure (for all values of $N$) of those algebras,
including the expression of generators themselves, in terms of matrices
with coefficients
in $\CC$ or $Gr(2)$, was obtained by  \cite{Oleg}.
Using these results, the representation theory of ${\mathcal H}$,
for the case $N=3$, was presented in \cite{Coquereaux}.
Following this work, the authors of \cite{Dabrowski}, studied
the action of ${\mathcal H}$ (case $N=3$) on the algebra of complex
matrices $M_3(\CC)$.
In the letter \cite{CoGaTr}, a reduced Wess-Zumino complex $\Omega_{WZ}({\mathcal
M})$ was introduced, thus providing a differential calculus bicovariant with respect 
to the action of the quantum group ${\mathcal H}$ on the algebra $M_3(\CC)$  
of complex matrices. This differential algebra (that could be used to generalize
gauge field theory models on an auxiliary smooth manifold) was also analysed
in terms of representation theory of ${\mathcal H}$ in the same letter. 
In particular, it was shown that $M_3(\CC)$ itself can be
reduced into the direct sum of three indecomposable representations
of ${\mathcal H}$.  A general discussion of several other properties of the dually paired 
Hopf algebras ${\mathcal F}$ and ${\mathcal H}$ (scalar products, star
structures, twisted derivations \etc) can also be found there, as well as in the article \cite{CGTrev}. 
Other properties  of $SL_{q}(2,\CC)$ at third (or fourth) root of unity 
should also be discussed in  \cite{Kastler-Valavane}.

In the present paper, after recalling several basic definitions and
properties, we show that the algebra of usual $N\times N$
complex matrices (assuming $N$ odd) decomposes, under the action of the
quantum group
${\mathcal H}$, into a direct sum of $N$ indecomposable representations
of dimension $N$ that we call ${\sl N}_{p}$. 
Every {\sl indecomposable\/} module of this type contains an invariant
{\sl irreducible\/} subspace of dimension $p$. In particular, ${\sl 
N}_{N}$ itself is irreducible.
We also show how these 
representations appear as particular subrepresentations of the projective 
indecomposable modules of ${\mathcal H}$.

Finally we shall give the action of generators of ${\mathcal H}$
on the elements of the reduced Wess-Zumino complex.

A short discussion about the use of those structures in physics is 
given at the end.

\section{The dually paired finite dimensional quantum groups ${\mathcal
F}$ and ${\mathcal H}$}

\subsection{${\mathcal M} \doteq M_N(\CC)$ as a finite dimensional
quantum plane}
The algebra of $N \times N$ matrices can be generated by two
elements $x$ and $y$ with relations :
\begin{equation}
   xy = q yx \qquad \text{and} \qquad x^N = y^N = \one \ ,
\label{M-relations}
\end{equation}
where $q$ denotes a $N$-th root of unity ($q \neq 1$) and $\one$ denotes
the unit matrix. Explicitly, $x$ and $y$ can be taken as the
following matrices:
\begin{equation}
x = \begin{pmatrix}  1 & & & & \cr & q^{-1} & & & \cr
 & & q^{-2} & & \cr   & & & \ddots & \cr & & & & q^{-(N-1)} \end{pmatrix}
\qquad   
y = \begin{pmatrix} 0 & & & & \cr \vdots & & & & \cr 
\vdots & & \one_{N-1} & & \cr \vdots & & & & \cr 1 & 0 & \cdots &\cdots & 0 
\end{pmatrix}
\label{xymatrix}
\end{equation}
This result can be found in \cite{Weyl}.\\
{ \bf Warning} : for technical reasons, we shall assume in all this paper
that $N$ is odd.

\subsection{The dually paired quantum groups ${\mathcal F}$  and ${\mathcal
H}$}
\subsubsection{The quantum group ${\mathcal F}$ coacts on ${\mathcal
M}$ }
We now consider a free associative algebra generated by four {\it a priori\/}
non commuting symbols $a,b,c,d$ and define the following ``change of
variables'' (notice that the symbol $\otimes$ does not denote the usual
tensor product
since it involves also a matrix multiplication):
\begin{equation}
\begin{pmatrix} x' \cr y' \end{pmatrix} =
    \begin{pmatrix} a & b \cr c & d \end{pmatrix} \otimes 
\begin{pmatrix} x \cr y \end{pmatrix} \ ,
\label{c-L-coaction}
\end{equation}
and
\begin{equation}
\begin{pmatrix} \tilde x & \tilde y \end{pmatrix}  =
    \begin{pmatrix} x & y \end{pmatrix} \otimes 
\begin{pmatrix} a & b \cr c & d \end{pmatrix} \ .
\label{c-R-coaction}
\end{equation}
One then imposes that quantities $x',y'$ (and $\tilde x, \tilde y$)
obtained by the
previous matrix equalities should satisfy the same relations as $x$ and
$y$ (the multiplication of these elements make perfect sense within the tensor
product of the corresponding algebras). One obtains in this way the
relations:
\begin{equation}
\begin{tabular}{ll}
   $qca = ac$ \qquad \qquad & $qdb = bd$                \\
   $qba = ab$               & $qdc = cd$                \\
   $cb = bc$                & $ad-da = (q-q^{-1})bc$ \ ,\\
\end{tabular}
\label{F-products}
\end{equation}
together with
\begin{equation}
\begin{tabular}{ll}
   $a^N = \one \ , $ & $ b^N = 0    $ \\
   $c^N = 0    \ , $ & $ d^N = \one $ \\
\end{tabular}.
\label{F-products-quotient}
\end{equation}
One also take the central element ${\mathcal D} \doteq da -q^{-1}bc = ad -
qbc $ to be equal to $\one$.
The algebra defined by $a,b,c,d$ and the above set of
relation will be called  $\mathcal F$. It is clearly a quotient
of $Fun(SL_q(2,\CC))$, the algebra of polynomial functions on the
quantum group $SL_q(2,\CC)$. Since $a^{N} = \one$, multiplying the
relation $ad = \one + qbc$ from the
left by $a^{N-1}$ leads to
\begin{equation}
   d = a^{N-1} (\one + qbc)
\label{d-element-in-F}
\end{equation}
so that $d$ can be eliminated.
The algebra $\mathcal F$ can
therefore be {\sl linearly \/} generated ---as a vector space--- by the
elements $a^\alpha b^\beta c^\gamma $ where indices $\alpha, \beta, \gamma$
run in the set $\{0,1,\cdots ,N-1\}$. We see that $\mathcal F$ is a 
{\sl finite dimensional\/} associative algebra, whose dimension is
$$
   \dim ({\mathcal F}) = N^3 \ .
$$

${\mathcal F}$ is not only an associative algebra but a Hopf algebra 
(\cf previously given references). The coproduct of generators
 can be read directly from the above $2\times 2$ matrix with entries 
 $a,b,c,d$, for instance, $\Delta a = a \otimes a + b \otimes c$, \etc.

This quantum group, by
construction, {\sl coacts} on ${\mathcal M}$.

\subsubsection{The quantum group ${\mathcal H}$ acts on ${\mathcal
M}$ }

Multiplication and comultiplication being interchanged by
duality, it is clear that the dual ${\mathcal H}$ of ${\mathcal F}$ is also
a quantum
group (of the same dimensionality). For compatibility reasons with
previous references, we choose the linear basis $K^\alpha X_{+}^\beta
X_{-}^\gamma $
in ${\mathcal H}$, where $K,X_{+}$ and $X_{-}$ are defined by duality
as follows:
\begin{equation}
\begin{tabular}{cccc}
   $<K,a> = q$   & $<K,b> = 0$   & $<K,c> = 0$   & $<K,d> = q^{-1}$ \\
   $<X_+,a> = 0$ & $<X_+,b> = 1$ & $<X_+,c> = 0$ & $<X_+,d> = 0$ \\
   $<X_-,a> = 0$ & $<X_-,b> = 0$ & $<X_-,c> = 1$ & $<X_-,d> = 0$ \\
\end{tabular}
\label{H-F-pairing}
\end{equation}
From multiplication and comultiplication in ${\mathcal F}$, one gets :
\begin{description}

\item[Multiplication:]
\begin{eqnarray}
   K X_{\pm}     &=& q^{\pm 2} X_{\pm} K                 \nonumber \\
   \left[ X_+ , X_- \right]
                 &=& {1 \over (q - q^{-1})} (K - K^{-1})           \\
   K^N           &=& \one                                \nonumber \\
   X_+^N = X_-^N &=& 0 \ .                               \nonumber
\label{H-products}
\end{eqnarray}

\item[Comultiplication:]
The comultiplication is an algebra morphism, \ie
$\Delta(XY) = \Delta X \, \Delta Y$. It is given by
\begin{eqnarray}
   \Delta X_+ & \doteq & X_+ \otimes \one + K \otimes X_+      \nonumber \\
   \Delta X_- & \doteq & X_- \otimes K^{-1} + \one \otimes X_-           \\
   \Delta K & \doteq & K \otimes K                             \nonumber \\
   \Delta K^{-1} & \doteq & K^{-1} \otimes K^{-1} \ .          \nonumber
\label{H-coproducts}
\end{eqnarray}

There is also an antipode and a counit but we shall not need them in
the sequel.

\end{description}

The quantum group ${\mathcal H}$ acts on itself (by left or right
multiplication), it acts also on its dual ${\mathcal F}$ from the
left or from the right, as follows:
\begin{equation}
\scal{y}{h^{L}\cro{f}}   = \scal{yh}{f} \qquad 
\scal{y}{h^{R}\cro{f}}   = \scal{hy}{f} \ ,
\end{equation}
where $\, y,h \in \mathcal{H},\,  f \in \mathcal{F}$. These actions can also
be expressed as :
\begin{equation}
h^{L} \cro{f} = \scal{ \id \otimes h }{\Delta f}_{(2)} \qquad
h^{R} \cro{f} = \scal{ \id \otimes h }{\Delta f}_{(1)} \ ,
\end{equation}
where the notation $\langle \, , \, \rangle_{(1)}$ means that we only pair 
the first
term in the tensor product (resp. the second).\\
Finally, ${\mathcal H}$ also {\sl acts\/} on the reduced quantum plane
${\mathcal M}$
(the algebra of $N \times N$- matrices) since its dual ${\mathcal F}$
{\sl coacts\/} on it. There are again two possibilities, left or
right, but we shall use the left action.

The left action of ${\mathcal H}$ on ${\mathcal M}$ is generally
defined  as follows. If we denote the right coaction of $\mathcal{F}$ on 
$\mathcal{M}$ as :
\begin{equation}
\delta_R \cro{m} = m_{(1)} \otimes f_{(2)}  \hspace{2cm} 
\text{for}  \,\,\, m \in \mathcal{M} , f \in \mathcal{F} \ ,
\end{equation}
then :
\begin{equation}
h^L \cro{m} =  m_{(1)} \scal{h}{f_{(2)}}  \hspace {2cm} h \in \mathcal{H} \ .
\end{equation}
The action of generators of ${\mathcal H}$ on generators of ${\mathcal M}$
is given by the following table.
\begin{equation}
\begin{array}{|c||c|c|c|}
\hline
\text{Left} & K  & X_+ & X_-   \\
\hline
\hline
x & qx & 0 & y  \\
\hline
y & q^{-1}y & x & 0  \\
\hline 
\end{array}
\end{equation}
Using the coproduct, one finds the expression of these generators 
 on an arbitrary element of ${\mathcal M}$
\begin{eqnarray}
K^L \cro{x^{r}y^{s}}    &=& q^{(r-s)}x^{r}y^{s}    \nonumber   \\
X_+^L \cro{x^{r}y^{s}}  &=& q^{r}(1+q^{-2}+\cdots+q^{-2(s-1)})x^{r+1}y^{s-1}
\nonumber  \\
                        &=& q^{r}(\frac{1-q^{-2s}}{1-q^{-2}})x^{r+1}y^{s-1} \\
X_-^L \cro{x^{r}y^{s}}   &=& q^{s}(1+q^{-2}+\cdots+q^{-2(r-1)})x^{r-1}y^{s+1}
\nonumber   \\
                        &=& q^{s}(\frac{1-q^{-2r}}{1-q^{-2}})x^{r-1}y^{s+1}
\nonumber
\label{actionofHonM}
\end{eqnarray}
with $1 < r,s < N$.

Remember that ${\mathcal M}$ itself is {\sl not\/} a quantum group,
but a module (a representation space) for the quantum group ${\mathcal
H}$. However, taking $h \in {\mathcal H}$, $m_{1},m_{2} \in {\mathcal 
M}$ and writing $\Delta h = \sum 
h_{1}\otimes h_{2}$, one can check the following supplementary compatibility 
condition $h[m_{1}m_{2}] = \sum h_{1}[m_{1}]h_{2}[m_{2}]$
between the module structure of ${\mathcal M}$, the algebra structure 
of ${\mathcal M}$ and the coproduct in ${\mathcal
H}$. This makes ${\mathcal M}$ a module algebra over ${\mathcal H}$.

In order to reduce this module into indecomposable modules, it is
necessary to know at least part of the representation theory of ${\mathcal H}$. This
is recalled in the next subsection.

\subsection{Representation theory of ${\mathcal H}$}
As already stated in the introduction, using a result by
 \cite{Alekseev}, the explicit structure (for all values of $N$) of those
algebras,
including the expression of generators $X_{\pm}, K$ themselves, in terms of
matrices with coefficients
in $\CC$ or in the Grassmann\footnote{Remember that $\theta_1^{2} = 
\theta_2^{2} = 0$ and that $\theta_1 \theta_2 = -\theta_2 \theta_1$}
 algebra  $Gr(2)$ with two generators $\theta_1,\theta_2$, was
obtained by  \cite{Oleg}.
We shall not need the general theory but only the following fact:
when $N$ is odd, ${\mathcal H}$ is isomorphic with the direct sum
\begin{equation}
\mathcal{H} = M_N \oplus \Mlo{N-1|1} \oplus \Mlo{N-2|2} \oplus \cdots
                   \cdots  \oplus \Mlo{\frac{N+1}{2}|\frac{N-1}{2}}  
\label{isomorphisme}
\end{equation}
where :
\begin{itemize}
\item[-] $M_N$ is a $N\times N$ complex matrix
\item[-] An element of the $\Mlo{N-1|1}$ block (space that we shall just call 
$M_{N-1|1}$) is of the following form :
\begin{equation}
\begin{pmatrix} 
\bullet & \bullet & \cdots & \bullet & \bullet & \circ \cr
\bullet & \bullet & \cdots & \bullet & \bullet & \circ \cr
\vdots  & \vdots  &        &         & \vdots  & \vdots\cr
\bullet & \bullet & \cdots & \bullet & \bullet & \circ \cr
\circ   & \circ   & \cdots & \circ   & \circ   & \bullet \end{pmatrix}
\end{equation}
We have introduced the following notation:\\
$\bullet$ is an even element of the ring $Gr(2)$ of Grassmann numbers with 
two generators, \ie of the kind :\\
$$
\bullet = \alpha + \beta \theta_1 \theta_2 , \qquad \alpha,\beta \in \CC.
$$
$\circ$ is an odd element of the ring $Gr(2)$ of Grassmann numbers with 
two generators, \ie of the kind :\\
$$
\circ = \gamma\theta_1 + \delta \theta_2 \qquad \gamma, \delta \in \CC 
$$
\item[-] An element of the $M_{N-2|2}$ block is of the kind :
\begin{equation}
\begin{pmatrix}
\bullet & \bullet & \cdots & \bullet & \circ & \circ \cr
\bullet & \bullet & \cdots & \bullet & \circ & \circ \cr
\vdots  & \vdots  &        &         & \vdots  & \vdots\cr
\bullet & \bullet & \cdots & \bullet & \circ & \circ \cr
\circ  & \circ  & \cdots & \circ  & \bullet & \bullet \cr
\circ  & \circ   & \cdots & \circ   & \bullet   & \bullet \end{pmatrix}
\end{equation}
\item[-] \etc
\end{itemize}
Notice that ${\mathcal H}$ is \underline{not} a semi-simple algebra :
its Jacobson radical ${\mathcal J}$ is obtained by selecting in equation \ref{isomorphisme} the 
matrices with elements proportionnal to Grassmann 
variables. The quotient ${\mathcal H} / {\mathcal J}$ 
is then semi-simple\ldots but no longer Hopf!

Projective indecomposable modules (PIM's, also called principal 
modules) for ${\mathcal H}$ are
directly given by the columns of the previous matrices.
\begin{itemize}
	\item[-]  From the $M_{N}$ block, one obtains $N$ equivalent 
                  irreducible representations of dimension $N$ that we shall 
                  denote $N_{irr}$.
	\item[-]  From the $M_{{N-p}\vert p}$ block (\underline{assume} $p<N-p$), one obtains
	\begin{itemize}
		\item  $(N-p)$ equivalent indecomposable projective modules
                       of dimension $2N$ that we shall denote $P_{N-p}$
                       with elements of the kind 
          \begin{equation}
             (\underbrace{\bullet \bullet \cdots \bullet}_{N-p}
              \underbrace{\circ \circ \cdots \circ}_{p}  )
          \end{equation}

		\item   $p$ equivalent indecomposable projective modules (also of
dimension $2N$) that we shall
denote $P_{p}$ with elements of the kind 
\begin{equation}
( \underbrace{\circ \circ \cdots \circ}_{N-p}
  \underbrace{\bullet \bullet \cdots \bullet}_{p}  )
\end{equation} 
	\end{itemize}
\end{itemize}

Other submodules can be found by restricting the range of
parameters appearing in the columns defining the PIM's and imposing
stability under multiplication by elements of ${\mathcal H}$. 
In this way, one can determine, for each PIM the lattice 
of its submodules.
For a given PIM of dimension $2N$ (with the exception of $N_{irr}$), one
finds totally ordered sublattices (displayed below) with exactly three non 
trivial terms : the radical (here, it is the  biggest non trivial 
submodule of a given PIM),  the 
socle (here it is the smallest non trivial submodule), and one 
``intermediate'' submodule of 
dimension exactly equal to $N$. However the definition of this last 
submodule (up to equivalence) depends on the choice of an arbitrary complex parameter 
$\lambda$, so that we have a chain of inclusions for every such parameter. The 
collection of all these sublattices fully determines the lattice 
structure of submodules of a given principal module.
Since we have two types of principal modules (besides the irreducible 
ones,  $N_{irr}$), we  
obtain explicitly the following two types of chains of inclusions 
\footnote{Here we label the submodules by  underlining their
dimension.}:

\begin{itemize} 
	\item[-] First type : submodules of $P_{N-p}$.
 \begin{diagram}
\udl{0} & \rInto & \udl{N-p} & \rInto & {\udl{N}}_{N-p} 
& \rInto & \udl{N+p} & \rInto & \udl{2N} = P_{N-p}
\end{diagram}
where $\rInto$ represent inclusion.

An element of the submodule ${\udl{N}}_{N-p}$ (which has dimension $N$) is of the kind :
\begin{equation}
{\udl{N}}_{N-p} = 
\begin{pmatrix}  \beta_1\theta_1\theta_2 & \beta_2\theta_1\theta_2 & \cdots &
        \beta_{N-p}\theta_1\theta_2 & \gamma_1\theta_{\lambda}  &
        \cdots & \gamma_{p}\theta_{\lambda} \end{pmatrix} 
\end{equation}
where :
$$ \theta_{\lambda} = \lambda_1\theta_1 + \lambda_2\theta_2  \qquad 
\lambda = \lambda_1 / \lambda_2  \in \CC P^1$$
Notice that the submodule ${\udl{N}}_{N-p}$ itself is the direct sum of an invariant  sub-module of
dimension $(N-p)$, and a vector subspace of dimension $p$. We shall denote this as 
follows :
$$
{\udl{N}}_{N-p} = \udl{N-p} \inplus p
$$ 
with
\begin{equation}
\udl{N-p} = 
\begin{pmatrix}  \beta_1\theta_1\theta_2 & \beta_2\theta_1\theta_2 & \cdots &
        \beta_{N-p}\theta_1\theta_2 & 0  &
        \cdots & 0 \end{pmatrix} 
\end{equation}

	\item[-] Second type : submodules of $P_{p}$.
\begin{diagram}
\udl{0} & \rInto & \udl{p} & \rInto & {\udl{N}}_{p} 
& \rInto & \udl{2N-p} & \rInto & \udl{2N}' = P_{p}
\end{diagram}
An element of the submodule ${\udl{N}}_{p}$ (which has dimension $N$) is of the kind :
\begin{equation}
{\udl{N}}_{p} = 
\begin{pmatrix}  \gamma_1\theta_{\lambda_1} & \gamma_2\theta_{\lambda_2} & 
         \cdots & \gamma_{N-p}\theta_{\lambda_{N-p}} & 
        \beta_1\theta_1\theta_2 & \beta_2\theta_1\theta_2 & \cdots &
        \beta_{p}\theta_1\theta_2 \end{pmatrix}
\end{equation}

Notice that the submodule ${\udl{N}}_{p}$ itself is the direct sum of an invariant  sub-module of
dimension $p$, and a vector subspace of dimension $N-p$
$$
{\udl{N}}_{p} = \udl{p} \inplus (N-p)
$$
with
\begin{equation}
\underline{p}  =
\begin{pmatrix}  0 & 0 &
         \cdots & 0 & 
        \beta_1\theta_1\theta_2 & \beta_2\theta_1\theta_2 & \cdots &
        \beta_{p}\theta_1\theta_2 \end{pmatrix}
\end{equation}

\end{itemize}

Notice that the quotient of a PIM by its own radical defines an
irreducible representation for the quantum group ${\mathcal H}$. The
irreducible representations for ${\mathcal H}$ are therefore of dimensions
$1,2,3\ldots N$. Warning : the projective cover of a given indecomposable module
appearing in one of the above sublattices is not
necessarily equal to the principal module that appears as the maximum element of the same
sublattice.

We have already noticed that each PIM contains indecomposable submodules $N_{p}$ of dimension
exactly equal to $N$, each such submodule containing itself one invariant irreducible subspace of
dimension $p$ (the precise definition involves the choice of a 
parameter $\lambda$). As we shall see, these submodules of dimension 
$N$ are exactly those that appear in the decomposition of the algebra of 
complex $N\times N$ matrices  into representations of ${\mathcal H}$.

As an example, let us explicitly describe the case $N=5$. Then, $dim 
{\mathcal H} = 5^{3} = 125$. We can write $5 = 5+0 = 4+1 = 3+2$, so 
that ${\mathcal H} = M_5 \oplus \Mlo{4|1} \oplus \Mlo{3|2}$ and 
 we have five principal modules, one is irreducible ($5_{irr}$, of 
dimension $5$), the others ($P_{4}, P_{1}, P_{3}, P_{2}$) 
are projective indecomposable and have the same dimension $10$. One 
can also write  ${\mathcal H} = 5\, \udl{5}_{irr} \oplus 4\, P_{4} + 1 
\, P_{1} + 3 \, P_{3} + 2\,  P_{2}$.

The  lattices of submodules read:

\begin{diagram}
\udl{0} & \rInto & \udl{4} & \rInto & {\udl{5}}_{4} 
& \rInto & \udl{6} & \rInto & \udl{10}' = P_{4}
\end{diagram}
\begin{diagram}
\udl{0} & \rInto & \udl{1} & \rInto & {\udl{5}}_{1} 
& \rInto & \udl{9} & \rInto & \udl{10}'' = P_{1}
\end{diagram}
\begin{diagram}
\udl{0} & \rInto & \udl{3} & \rInto & {\udl{5}}_{3} 
& \rInto & \udl{7} & \rInto & \udl{10}''' = P_{3}
\end{diagram}
\begin{diagram}
\udl{0} & \rInto & \udl{2} & \rInto & {\udl{5}}_{2} 
& \rInto & \udl{8} & \rInto & \udl{10}'''' = P_{2}
\end{diagram}

Besides the five-dimensional irreducible representation which is 
itself a PIM, there are four others : $\udl{4}_{irr}= P_{4}/\udl{6}$, 
$\udl{1}_{irr}= P_{1}/\udl{9}$, $\udl{3}_{irr}= P_{3}/\udl{7}$ and $\udl{2}_{irr}= 
P_{2}/\udl{8}$. 

\section{Reduction of the algebra ${\mathcal M} \doteq M_N(\CC)$, $N$ odd, into indecomposable representations of ${\mathcal H}$}

We shall now focus our attention on the case of $N\times N$ complex 
matrices, in the case where $N$ is odd, and show how this familiar algebra can 
be reconstructed as a sum of representations for the finite 
dimensional quantum 
group ${\mathcal H}$.

 A vectorial basis of this 
algebra is given by matrices 
\begin{equation*}
\{ x^{r}y^{s} \}  \hspace{2cm} r,s \in 
\{0,1,\cdots,\text{N}-1  \}
\end{equation*}
where $x$ and $y$ are the particular generators already defined in 
section 2.
The left action of generators of $\mathcal{H}$ on an arbitrary 
element of $\mathcal{M}$ is given by:
\begin{eqnarray*}
K \cro{x^{r}y^{s}} &=& q^{(r-s)}x^ry^s  \\
X_+ \cro{x^ry^s}   &=& q^{r} \frac{1-q^{-2s}}{1-q^{-2}} x^{r+1}y^{s-1}  \\
X_- \cro{x^ry^s}   &=& q^{s} \frac{1-q^{-2r}}{1-q^{-2}} x^{r-1}y^{s+1}
\end{eqnarray*}
The generator $K$ always acts as an automorphism, for this reason, in order 
to study the invariant subspaces of   
$\mathcal{M}$ under the left action of $\mathcal{H}$, we shall only 
have to consider the action of $X_+$ et $X_-$.\\
Forgetting numerical factors, the action de $X_+$ and of 
$X_-$ on a given element of $\mathcal{M}$ can be written as follows:  
\begin{diagram}
x^{r+1}y^{s-1} & & \pile{\lTo^{X_+}  \\ \rDotsto_{X_-}} & & x^ry^s & & 
\pile{\lTo^{X_+}  \\ \rDotsto_{X_-}} &  & x^{r-1}y^{s+1}
\end{diagram}
It is then easy to decompose the space of $N\times N$ matrices into invariant
subspaces for this action.\\
$\bullet$ Starting from the element $x^{N-1}$, we have :
\begin{diagram}
0 & & & & & & & & & &  0 \\
\uTo^{X_+}  & & & & & & & & & &  \uDotsto ^{X_-} \\
 x^{N-1} & & \pile{\lTo^{X_+} \\ \rDotsto_{X_-}} & & x^{N-2}y & 
\cdots \cdots & xy^{N-2} & & \pile{\lTo^{X_+} \\ \rDotsto_{X_-}} & & 
y^{N-1} 
\end{diagram}
We obtain in this way an {\bfseries irreducible} subspace of 
dimension $N$, denoted
$N_{irr}$.
A base of this subspace is given by:
\begin{equation*}
\{ x^ry^s \}  \qquad \text{avec} \quad (r+s)=N-1 
\end{equation*} 
$\bullet$ Starting from the element $x^{N-2}$, we obtain the following 
diagram:
\begin{diagram}
0 & & & & & & & & & &  0 \\
\uTo^{X_+}  & & & & & & & & & &  \uDotsto ^{X_-} \\
 x^{N-2} & & \pile{\lTo^{X_+} \\ \rDotsto_{X_-}} & & x^{N-3}y & 
\cdots \cdots \cdots & xy^{N-3} & & \pile{\lTo^{X_+} \\ \rDotsto_{X_-}} & & 
\,\,y^{N-2}  \\
 & \luDotsto(4,2)_{X_-} & & & & & & & &\ruTo(4,2)_{X_+} &  \\
 & & & & & x^{N-1}y^{N-1} & & & & &    
\end{diagram}
We obtain again an invariant subspace of the same dimension $N$, 
denoted ${N}_{N-1}$. 
A base of this subspace is given by:
\begin{equation*}
\{ x^ry^s \}  \qquad \text{avec} \quad (r+s)=N-2 \quad [\text{modulo}\,N] 
\end{equation*}
This space is the direct sum of an invariant subspace of dimension 
$N-1$ (hence the notation), a basis of which being given by
\begin{equation*}
\{ x^ry^s \}  \qquad \text{avec} \quad (r+s)=N-2 ,
\end{equation*}
and a non-invariant vector subspace of dimension $1$ generated by  
$\{ x^{N-1}y^{N-1} \}$.\\

$\bullet$ The process can be repeated, up to:
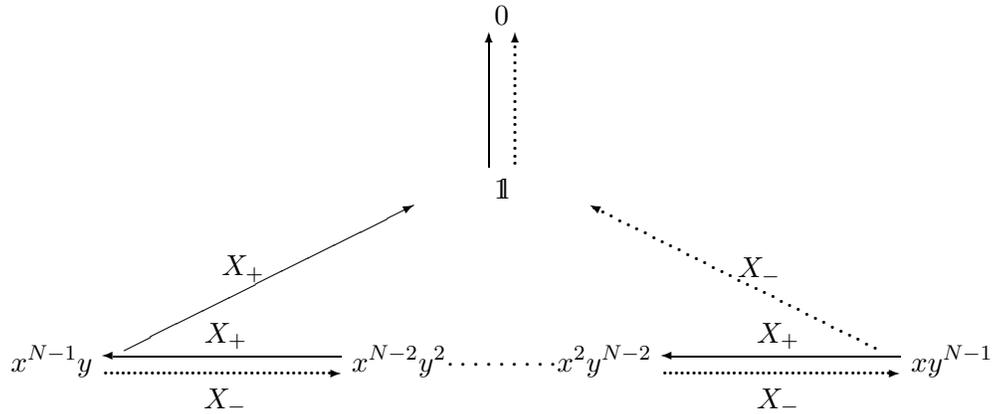
\begin{diagram}
 & & & & &            0                  & & & & & \\
 & & & & &   \uTo  \uDotsto              & & & & & \\
 & & & & &         \one               & & & & & \\  
 & & & &\ruTo(5,2)^{X_+}  & & \luDotsto(5,2)^{X_-}& & & & \\
x^{N-1}y & &  \pile{\lTo^{X_+} \\ \rDotsto_{X_-}} &  & x^{N-2}y^2 & 
\cdots \cdots \cdots & x^{2}y^{N-2} & &  \pile{\lTo^{X_+} \\ \rDotsto_{X_-}}
 & & xy^{N-1}
\end{diagram}
We obtain in this last case a subspace of still the same dimension  
$N$, denoted  
${N}_{1}$ with a base given by:
\begin{equation*}
\{ x^ry^s \}  \qquad \text{avec} \quad (r+s)=0 \quad [\text{modulo}\,N]
\end{equation*}
It is the direct sum of an invariant vector subspace of dimension 
$1$, generated by $\one$, and a non-invariant vector subspace of 
dimension $N-1$.

To conclude, we see that, under the left action of $\mathcal{H}$, the 
algebra of $N\times N$ matrices can be decomposed into a direct sum 
of invariant subspaces of dimension $N$, according to
\begin{itemize}
\item $N_N = N_{irr}$ : irreducible
\item $N_{N-1}$ : reducible indecomposable, with an invariant subspace 
      of dimension $N-1$.
\item $N_{N-2}$ : reducible indecomposable, with an invariant subspace 
      of dimension $N-2$.
\item $\vdots$
\item $N_{1}$ : reducible indecomposable, with an invariant subspace 
of dimension $1$.
\end{itemize}
These representations of dimension 
$N$ coincide exactly with those also called $N_{p}$ (or $N_{N-p}$) in the previous section.
Using these notations, the algebra of matrices
N$\times$N can be written
\begin{equation}
\mathcal{M} = N_N \oplus N_1 \oplus N_2 \oplus \cdots \oplus N_{N-1}
\end{equation}
with:
\begin{equation*}
\begin{array}{ccrcl}
N_N &=& N_{irr} & &  \\
N_{N-1} &=& (N-1) &\inplus& 1 \\
N_{N-2} &=& (N-2) &\inplus& 2 \\
\vdots  \\
N_{1} &=& 1 &\inplus& (N-1)
\end{array}
\end{equation*} 

Continuing the example given at the end of section $2$ we see that 
$$M(5,\CC) = \underline{5} \oplus \underline{5}_{4} \oplus \underline{5}_{3} \oplus 
\underline{5}_{2} \oplus \underline{5}_{1}$$

\section{Action of ${\mathcal H}$ on the reduced Wess-Zumino complex $\Omega_{WZ}({\mathcal M})$}

\subsection{The reduced Wess-Zumino complex $\Omega_{WZ}({\mathcal M})$}

The Wess-Zumino complex was constructed in \cite{Wess} as the unique 
(up to a redefinition of $q \rightarrow q^{-1}$) quadratic differential algebra on the 
quantum plane, 
bicovariant with respect to the action of the quantum group 
$SL_{q}(2,\CC)$. 
The commutation relations between the $N \times N$ matrices $x$, $y$ and their 
differentials $dx$, $dy$, and between the differentials themselves are given by:

\begin{equation}
\begin{tabular}{ll}
   $xy = qyx$                 &                                        \\
   $x\,dx = q^2 dx\,x$ \qquad & $x\,dy = q \, dy\,x + (q^2 - 1) dx\,y$ \\
   $y\,dx = q \, dx\,y$       & $y\,dy = q^2 dy\,y$                    \\
   $dx^2 = 0$                 & $dy^2 = 0$                             \\
   $dx\,dy + q^2 dy\,dx = 0$  &                                        \\
\end{tabular}
\label{WZ-relations}
\end{equation}

The reduced quantum plane itself is not a quadratic algebra, 
since it contains relations like $x^{N}=y^{N}=1$, but Leibniz rule, 
together with the fact that $N$ is a root of unity imply that such 
compatibility relations like $dx^{N}=dy^{N}=0$ are automatically satisfied 
\cite{CoGaTr}\cite{CGTrev}, indeed

\begin{eqnarray*}
   d(x^N) &=& d(x^{N-1}) x + x^{N-1} dx = \ldots =
              (1+q+q^2+\ldots + q^{N-1}) x^{N-1} dx = 0 \\
   d(y^N) &=& d(y^{N-1}) y + y^{N-1} dy = \ldots  =
               (1+q+q^2+\ldots + q^{N-1}) x^{N-1} dy= 0
\end{eqnarray*}

We call ``reduced Wess-Zumino complex'' $\Omega_{WZ}({\mathcal M})$ the quotient of the 
differential algebra of Wess-Zumino by the corresponding differential 
ideal. 
Note that $\dim(\Omega_{WZ}^0({\mathcal M})) = N^{2}$, $\dim(\Omega_{WZ}^1({\mathcal M})) = 2 N^{2}$
and $\dim(\Omega_{WZ}^2({\mathcal M})) = N^{2}$. Therefore, 
$\dim(\Omega_{WZ}({\mathcal M}) = 4 
N^{2}$.

\subsection{Remarks}
Here is a list of questions that can be asked about $\Omega_{WZ}({\mathcal 
M}) $.

\begin{itemize}
	\item  Study the action of ${\mathcal H}$.

	\item  Decompose this differential algebra into representations of ${\mathcal H}$

	\item  Study the cohomology of $d$

	\item  Extend the star operation(s) of ${\mathcal M}$ to star (or graded 
star) operations of the differential algebra

	\item  Study algebraic connections in $\Omega_{WZ}({\mathcal M}) $ and their 
curvature
\item \etc
\end{itemize}

These questions are studied in \cite{CoGaTr}\cite{CGTrev}, mostly with the 
particular choice $N=3$. Here we shall only give explicitly the action 
of  ${\mathcal H}$ on this differential algebra, for an arbitrary $N$.

\subsection{Action of ${\mathcal H}$}

We already gave in \ref{actionofHonM} the left action of generators of $\mathcal{H}$ 
on an arbitrary element of $\mathcal{M}$.
The left action of the same elements on generators $dx,dy$ of the Manin dual 
$\mathcal{M}^{!}$ of $\mathcal{M}$ (although the later is not quadratic, we 
can use this terminology !) is given by :

\begin{xalignat*}{3}
K^{L} \cro{dx} &= qdx &\qquad X_+^{L} \cro{dx} &=0 &
                  \qquad X_-^{L}\cro{dx}&=dy\\
K^{L} \cro{dy} &= q^{-1}dy &\qquad X_+^{L} \cro{dy} &=dx 
	         &\qquad X_-^{L}\cro{dy}&=0
\end{xalignat*}

Using these two pieces of information, we now find the action of generators of $\mathcal{H}$ on arbitrary elements of 
$\Omega_{WZ}^{1}(\mathcal{M})$ thanks to the coproduct :  \\
Let us write : $\Delta (h)=h_{(1)} \otimes h_{(2)}$ whenever $h \in \mathcal{H} $, 
then, 
for  $m_1 \in \mathcal{M}$ and $m_2 \in \mathcal{M}^{!}$, we have:
\begin{equation*}
h^{L} \cro{m_1m_2} = h_{(1)}^{L} \cro{m_1} \otimes h_{(2)}^{L} \cro{m_2} 
\end{equation*}

The left action of generators of $\mathcal{H}$ on arbitrary elements 
of $\Omega_{WZ}^{1}(\mathcal{M})$ is then given by :

\begin{eqnarray*}
K   \cro{x^ry^sdx} &=& q^{r+1-s}x^ry^s dx \\
K   \cro{x^ry^sdy} &=& q^{r-s-1}x^ry^s dy \\
X_+ \cro{x^ry^sdx} &=& q^r \frac{1-q^{-2s}}{1-q^{-2}}x^{r+1}y^{s-1} dx  \\
X_+ \cro{x^ry^sdy} &=& q^r \frac{1-q^{-2s}}{1-q^{-2}}x^{r+1}y^{s-1} dy
                    + q^{r-s} x^r y^s dx  \\   
X_- \cro{x^ry^sdx} &=& q^{s-1} \frac{1-q^{-2r}}{1-q^{-2}}x^{r-1}y^{s+1} dx
                    +  x^r y^s dy  \\
X_- \cro{x^ry^sdy} &=& q^{s+1} \frac{1-q^{-2r}}{1-q^{-2}}x^{r-1}y^{s+1} dx  
\end{eqnarray*}

\section{Comments}

The usual algebra ${\mathcal M}$ of $N \times N$ complex matrices
(or the group of its unitary elements) is very often 
used in various fields of physics. The observation that it is a module 
algebra for a finite dimensional non commutative, non 
cocommutative (and non semi-simple)
quantum group, and that it can be correspondingly decomposed into the 
sum of $N$ indecomposable representations (of dimension $N$)
came up --- for us --- as a surprise. One could then be tempted to speak 
of ``hidden symmetry'', whenever  ${\mathcal M}$ plays a role in 
the description of some physical process, 
but we do not know yet the physical meaning of such a ``symmetry'', 
and its interpretation should certainly depend upon the particular 
situation at hand. The above properties nevertheless suggest
that, in several branches of physics, it may be worth to study the 
appearance and meaning of ``symmetries'' associated with this
 quantum group ${\mathcal H}$ (which is a kind of ``fat'' version of the 
algebra of the group $\ZZ_{N}$).

Notice that representations of ${\mathcal H}$ are also, {\it a priori\/}, 
particular representations of $U_{q}sl(2,\CC)$ when $q$ is a root of 
unity (representations for which 
$K^{N}=1$, and $X_{\pm}^{N}=0$). Such representations appear in 
several examples of conformal field theories, and also in the study 
of abelian anyons (particles obeying a one-dimensional statistics associated 
with the braid group, or with a Hecke algebra). It may be simpler 
and conceptually more appropriate to discuss such problems in terms 
of ${\mathcal H}$ than with the infinite dimensional Hopf algebra $U_{q}sl(2,\CC)$.

One of the purposes of the letter \cite{CoGaTr} was to construct 
generalized  differential forms as elements of the tensor product  $\Xi$
of the space of usual (De Rham) forms on a (usual) space-time and of the differential algebra 
defined by the reduced Wess-Zumino complex. One obtains naturally on 
$\Xi$ an action of a Hopf algebra which is the product of ${ \mathcal H}$, 
times the envelopping algebra of the Lie algebra of the
Lorentz group. ${\mathcal H}$ appears therefore as a discrete (but 
neither commutative nor cocommutative) analogue of the Lorentz group.

Finally, the group of unitary elements of the quotient of ${\mathcal H}$ 
by its radical, when $N$ is odd, is isomorphic with
$$U(N)\times (U(N-1) \times U(1)) \times (U(N-2) \times U(2)) \times (U(N-3) 
\times U(3)) \times \ldots$$
For $N = 3$, one obtains $U(3)\times U(2) \times U(1)$. After appropriate 
identification of several $U(1)$ factors, one recognizes the gauge group 
describing of the standard model of elementary particles. This last observation was made in
\cite{Connes} where it was suggested that ${\mathcal H}$ could 
play the role of a Lorentz group for the ``internal space'' and that 
it would be tempting to devise a generalized gauge theory using this 
finite quantum group as a basic ingredient. 
This last comment can also be related to the construction described in  \cite{CoGaTr}.


\begin{thebibliography}{99}

\bibitem{Dabrowski} L. D{\c{a}}browski, F. Nesti and P. Siniscalco,
   {\em A finite quantum symmetry of $M(3,\CC)$}, SISSA 63/97/FM,
   {\tt hep-th/9705204}.

\bibitem{Gluschenkov} D. V. Gluschenkov and A. V. Lyakhovskaya,
   {\em Regular representation of the quantum Heisenberg double ($q$ is a
   root of unity)}, Zapiski LOMI 215 (1994).

\bibitem{Alekseev} A. Alekseev, D. Gluschenkov and A. Lyakhovskaya,
   {\em Regular representation of the quantum group $SL_q(2)$ ($q$ is a root
   of unity)}, St. Petersburg Math. J., Vol {\bf 6}, N5, 88 (1994).

\bibitem{Coquereaux} R. Coquereaux, {\em On the finite dimensional quantum
   group $M_3 \oplus (M_{2\vert 1}(\Lambda^2))_0$}, CPT-96/P.3388,
   {\tt hep-th/9610114}, Lett. Math. Phys. {\bf 42}, 309 (1997).

\bibitem{Wess} J. Wess and B. Zumino,
   {\em Covariant differential calculus on the quantum hyperplane},
   Nucl. Phys. B (Proc. Suppl.) {\bf 18B}, 302 (1990).

\bibitem{Weyl} H. Weyl, {\em The theory of groups and quantum mechanics},
   Dover Publications (1931).

\bibitem{Oleg} O. Ogievetsky, {\em Matrix structure of $SL_q(2)$
   when $q$ is a root of unity}, CPT-96/P.3390, to appear.

\bibitem{CoGaTr} R. Coquereaux, A. O. Garc\'{\i}a and R. Trinchero,
   {\em Finite dimensional quantum group covariant differential calculus
   on a complex matrix algebra}, CPT-98/P.3630, {\tt math.QA/9804021}.
   
\bibitem{CGTrev} R. Coquereaux, A. O. Garc\'{\i}a and R. Trinchero,
   {\em Differential calculus and connections on a quantum plane at a 
   cubic root of unity}, CPT-98/P.3632, {\tt math-ph/9807012}


\bibitem{Connes} A. Connes, {\em NonCommutative Geometry and Reality},
   IHES/M/95/52.
   
\bibitem{Kastler-Valavane} D. Kastler, Valavane, {\em } 
Proceedings of the conference of Palermo (1997), to appear.   

\end{thebibliography}
\end{document}